\documentclass[dvipdfmx]{jpsj-suppl}
\usepackage{txfonts} 

\def\mev{\text{ MeV}}
\def\gev{\text{ GeV}}

\def\microB{\text{ }\mu\text{b}}
\def\HeT{{}^{3} \text{He}}

\title{The Peak Structure in the In-Flight ${}^{3}\text{He} ( K^{-} ,
  \, \Lambda p ) n$ Reaction Around the $\bar{K} N N$ Threshold}

\author{Takayasu \textsc{Sekihara}$^{1}$, Eulogio \textsc{Oset}$^{2}$,
  and Angels \textsc{Ramos}$^{3}$}

\inst{$^{1}$Advanced Science Research Center, Japan Atomic
    Energy Agency, Shirakata, Tokai, Ibaraki, 319-1195, Japan \\
    $^{2}$Departamento de F\'{\i}sica Te\'orica and IFIC, Centro
    Mixto Universidad de Valencia-CSIC, Institutos de Investigaci\'on
    de Paterna, Aptdo. 22085, 46071 Valencia, Spain \\
    $^{3}$Departament de F\'{\i}sica Qu\`antica i Astrof\'{\i}sica
    and Institut de Ci\`encies del Cosmos, Universitat de Barcelona,
    Mart\'i i Franqu\`es 1, 08028 Barcelona, Spain}

\email{sekihara@post.j-parc.jp}

\recdate{October 16, 2016}

\abst{We theoretically investigate the origin of the peak structure
  around the $K^{-} p p$ threshold observed in the in-flight
  ${}^{3}\text{He} ( K^{-} , \, \Lambda p ) n$ reaction in the recent
  E15 experiment at J-PARC.  For this purpose, we consider two
  scenarios to produce the peak.  One is that the $\Lambda (1405)$ is
  generated but it does not correlate with $p$, and the uncorrelated
  $\Lambda (1405) p$ system subsequently decays into $\Lambda p$. The
  other one is that the $\bar{K} N N$ bound state is indeed generated and
  decays into $\Lambda p$.  As a result, the experimental signal is
  qualitatively well reproduced in the $\bar{K} N N$ bound scenario,
  definitely discarding the uncorrelated $\Lambda (1405) p$ one.}

\kword{kaonic nuclei, J-PARC E15 experiment, chiral dynamics}

\begin{document}
\maketitle

\section{Introduction}

Because the interaction between antikaon ($\bar{K}$) and nucleon ($N$)
is strongly attractive\cite{Kaiser:1995eg, Oset:1997it}, we expect
that there should exist bound states of $\bar{K}$ and nuclei.  They
are referred to as kaonic nuclei.  There are at least two motivations
to study kaonic nuclei.  One is that kaonic nuclei are exotic states
of many-body systems tied by the strong interaction, which contain a
hadron other than nucleons, and the other is that kaonic nuclei are a
good ``laboratory'' to prove kaons in finite nuclear density.  In
particular, the simplest kaonic nucleus $\bar{K} N N ( I = 1/2 )$,
sometimes called $K^{-} p p$, has been intensively studied in both
theoretical\cite{Akaishi:2002bg, Shevchenko:2006xy, Ikeda:2007nz,
  Dote:2008in, Wycech:2008wf, Bayar:2011qj, Barnea:2012qa} and
experimental\cite{Agnello:2005qj, Yamazaki:2010mu, Tokiyasu:2013mwa,
  Ichikawa:2014ydh, Hashimoto:2014cri} sides (see also
Ref.\cite{Nagae:2016cbm}).  However, at present, there has been no
consensus on the properties of the $\bar{K} N N$ bound state.

In this line, a very attractive result comes out from the recent
J-PARC E15 experiment\cite{Sada:2016vkt}.  In this experiment, they
observed a peak structure near the $K^{-} p p$ threshold in the
$\Lambda p$ invariant mass spectrum of the in-flight $\HeT (K^{-} ,
\Lambda p) n$ reaction with a kaon momentum in the laboratory frame of
$k_{\rm lab} = 1 \gev /c$.  This peak can be described by the
Breit-Wigner formula with mass $M_{X} = 2355 ^{+6}_{-8} \text{(stat.)}
\pm 12 \text{(sys.)} \mev$ and width $\Gamma _{X} = 110^{+19}_{-17}
\text{(stat.)} \pm 27 \text{(sys.)} \mev$\cite{Sada:2016vkt}.  This
could be a signal of the $\bar{K} N N ( I = 1/2 )$ bound state with a
binding $\sim 15 \mev$ from the $K^{-} p p$ threshold.

In this study, our motivation is to investigate what is the origin of
the peak structure observed in the J-PARC E15
experiment\cite{Sada:2016vkt}.  For this purpose we take into account
two possible mechanisms for producing a peak in the mass spectrum.
One is that the $\Lambda (1405)$ is generated but it does not
correlate with the $p$, and the uncorrelated $\Lambda (1405) p$ system
subsequently decays into $\Lambda p$.  This uncorrelated $\Lambda
(1405) p$ system may create a peak even if they do not bind because
the $\Lambda (1405)$ exists below the $\bar{K} N$ threshold.  The
other is that the $\bar{K} N N$ bound state is indeed generated and
decays into $\Lambda p$, which can create a peak in the $\Lambda p$
invariant mass spectrum if the signal is strong enough.  In the
following, we concentrate on the ${}^{3}\text{He} ( K^{-} , \, \Lambda
p ) n$ reaction with a kaon momentum in the laboratory frame of
$k_{\rm lab} = 1 \gev /c$.  The details of the calculations are given
in Ref.\cite{Sekihara:2016vyd}.

\section{Uncorrelated $\Lambda (1405) p$ scenario}

\begin{figure}[t]
  \centering
  \begin{minipage}{0.35\hsize}
    \includegraphics[scale=0.169]{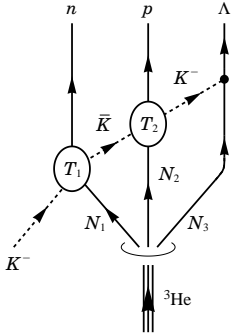}
  \end{minipage}
  \begin{minipage}{0.55\hsize}
    \caption{Feynman diagram most relevant to the three-nucleon
      absorption of a $K^{-}$ via an uncorrelated $\Lambda (1405) p$
      system\cite{Sekihara:2016vyd}.  We take into account the
      antisymmetrization for the three nucleons in $\HeT$.}
    \label{f1}
  \end{minipage}
\end{figure}

First of all, we would like to consider the scenario of the
uncorrelated $\Lambda (1405)$ and proton.  For this scenario we employ
the most relevant diagram shown in Fig.~\ref{f1}.  The scattering
amplitude of this reaction is fixed as follows.  The $\HeT$ wave
function is obtained as an antisymmetrized system of three nucleons in
a harmonic oscillator potential.  The amplitudes of the first
collision, $T_{1} ( K^{-} n \to K^{-} n)$ and $T_{1} ( K^{-} p \to
\bar{K}^{0} n)$, are taken from the differential cross sections of
each reaction at $k_{\rm lab} = 1 \gev /c$.  The amplitudes of the
second collision, $T_{2} ( K^{-} p \to K^{-} p)$ and $T_{2} (
\bar{K}^{0} n \to K^{-} p)$, are calculated in the chiral unitary
approach\cite{Kaiser:1995eg, Oset:1997it, Oller:2000fj, Jido:2003cb}
and we effectively take into account the kaon absorption width in the
kaon propagator.  The $K^{-} p \Lambda$ vertex is fixed by chiral
perturbation theory, and the intermediate kaon energy is determined in
two options: the Watson approach (A) and truncated Faddeev approach
(B)~\cite{Jido:2012cy}.

\begin{figure}[b]
  \centering
  \includegraphics[width=7.5cm]{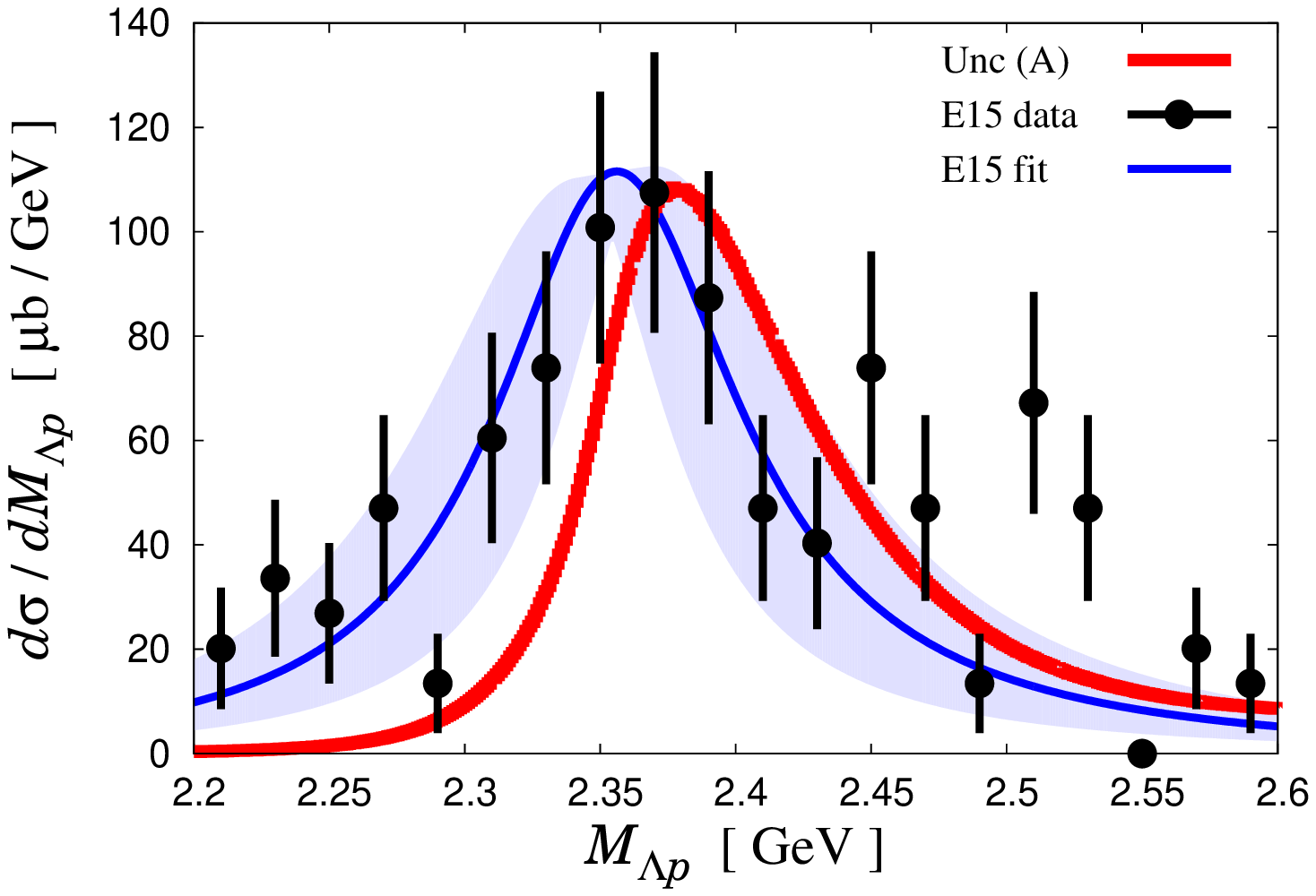} ~
  \includegraphics[width=7.5cm]{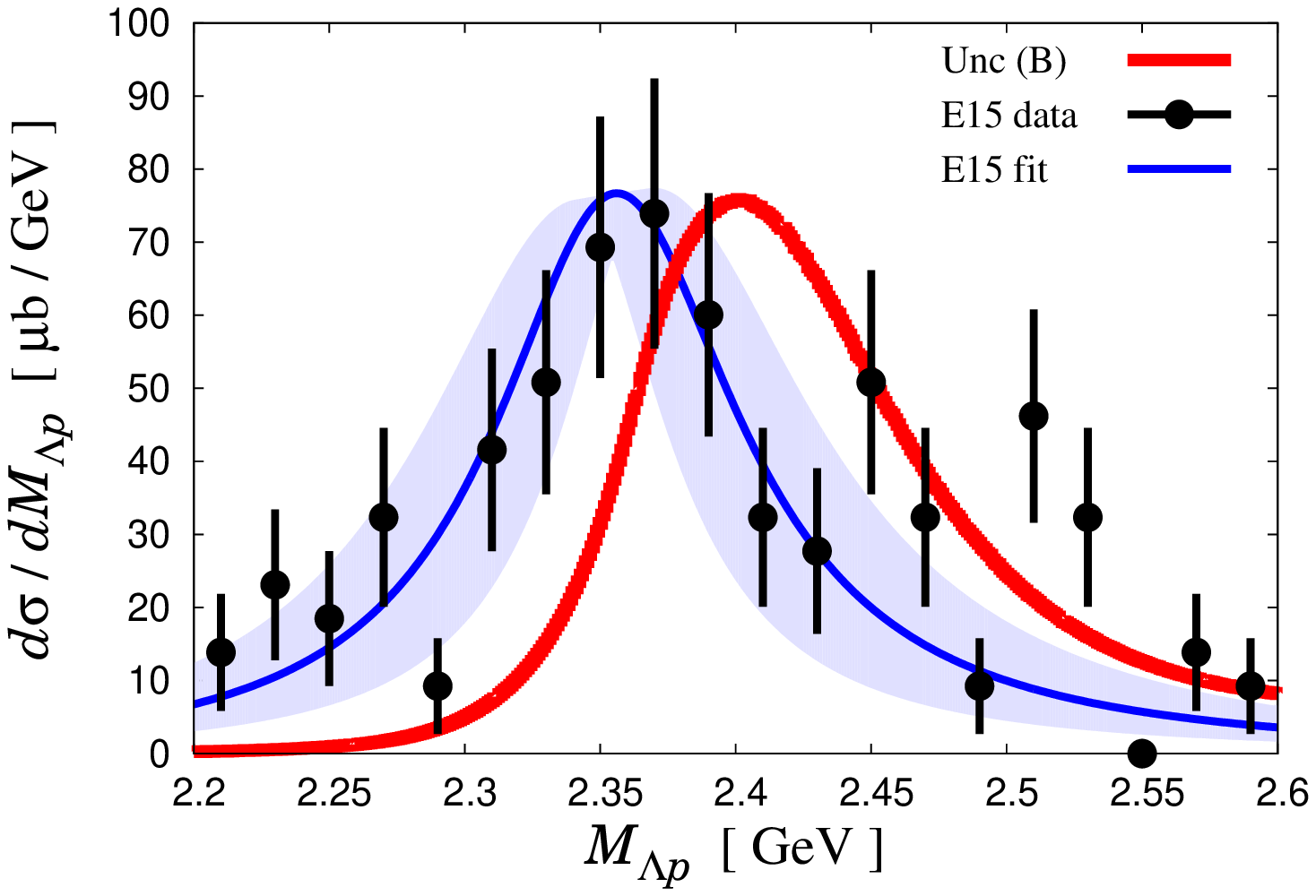}
  \caption{Mass spectrum for the $\Lambda p$ invariant mass of the
    in-flight ${}^{3}\text{He} ( K^{-} , \, \Lambda p ) n$ reaction in
    the uncorrelated $\Lambda (1405) p$ scenario.  Calculations are
    done in option A (left) and option B
    (right)\cite{Sekihara:2016vyd}.  The experimental (E15) data and
    its fit are taken from Ref.\cite{Sada:2016vkt} and shown in
    arbitrary units.}
\label{f2}
\end{figure}

Now we can calculate the $\Lambda p$ invariant mass spectrum of the
reaction ${}^{3}\text{He} ( K^{-} , \, \Lambda p )
n$\cite{Sekihara:2016vyd}, which is shown in Fig.~\ref{f2} together
with the experimental (E15) data and its fit\cite{Sada:2016vkt} in
arbitrary units.  As one can see, in both options A and B, the peak
structure is inconsistent with the experimental data.  Actually, while
the peak appears at $2355 \mev$ in the experiment, we obtain the peak
at more than $2370 \mev$.  In particular, we cannot reproduce the
behavior of the lower tail $\sim 2.3 \gev$.  Therefore, the E15 signal
in the $\HeT (K^{-}, \Lambda p) n$ reaction is not the uncorrelated
$\Lambda (1405) p$ state.

\section{$\bar{K} N N$ bound scenario}

\begin{figure}[t]
  \centering
  \includegraphics[scale=0.169]{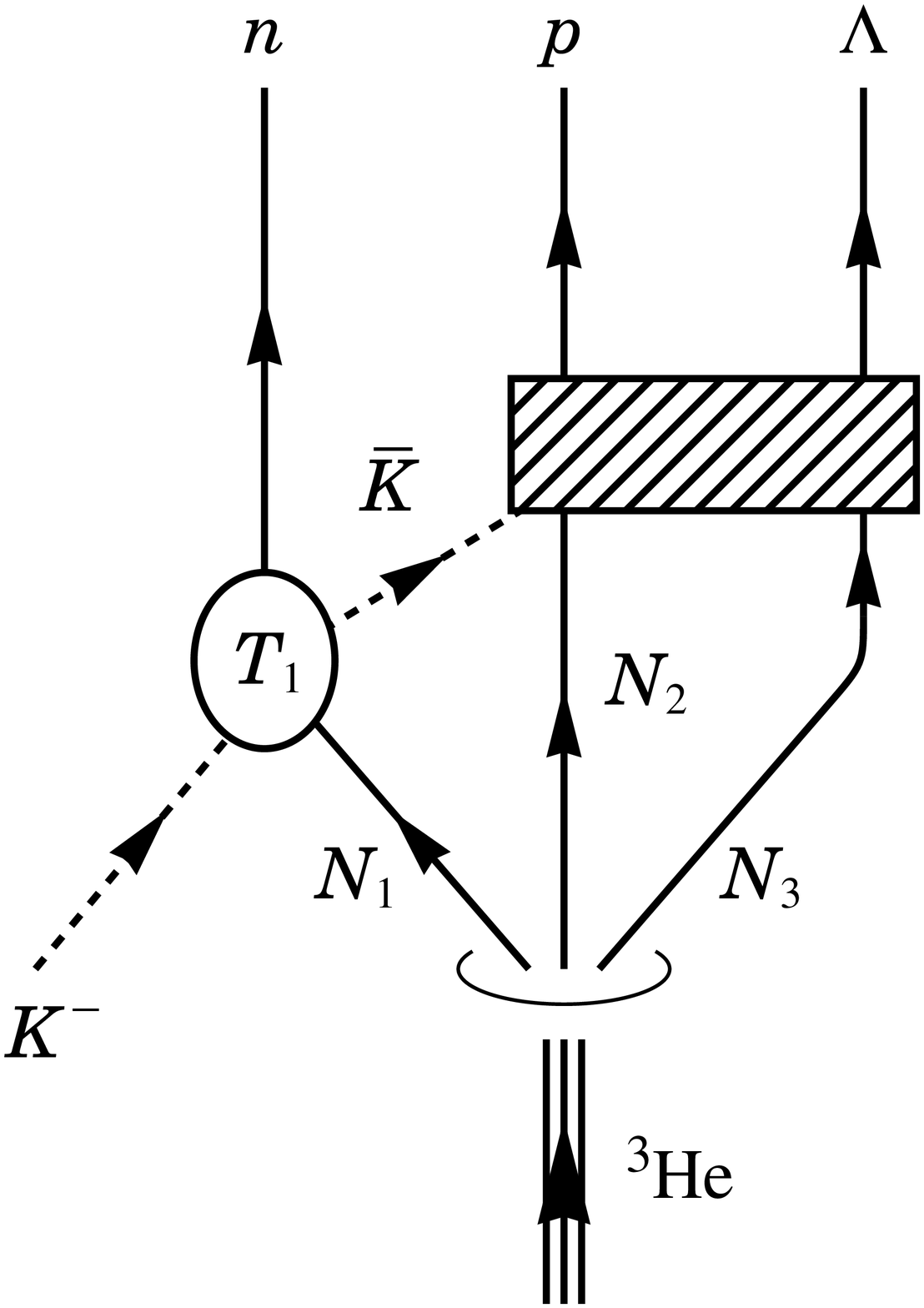} ~
  \includegraphics[scale=0.169]{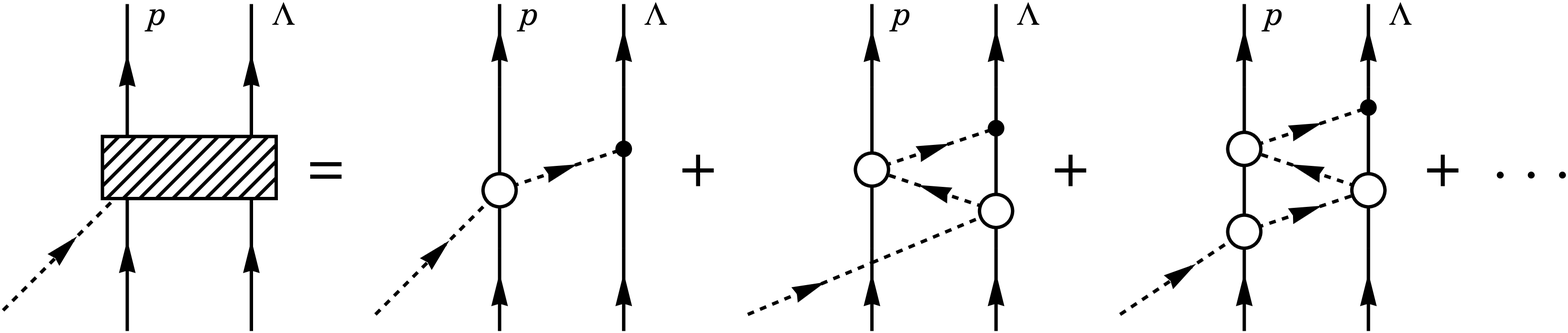}
  \caption{Feynman diagram most relevant to the three-nucleon
    absorption of a $K^{-}$ in the $\bar{K} N N$ bound scenario.  The
    multiple kaon scattering between two nucleons is shown as the
    diagrammatic equation in right\cite{Sekihara:2016vyd}, where
    dashed lines and open circles represent the kaon and the $\bar{K}
    N \to \bar{K} N$ amplitude in the chiral unitary approach,
    respectively.  We take into account the antisymmetrization for the
    three nucleons in $\HeT$.}
\label{f3}
\end{figure}

Next we consider the $\bar{K} N N$ bound scenario.  For this scenario
we employ the most relevant diagram shown in Fig.~\ref{f3}.  The
scattering amplitude of this reaction is fixed as follows.  We use the
same $\HeT$ wave function, amplitudes $T_{1}$, intermediate kaon
energy, and $\bar{K} N \Lambda$ vertex as in the uncorrelated $\Lambda
(1405) p$ case.  On the other hand, after the first collision of the kaon,
we make the multiple scattering of the kaon with two nucleons, which
is diagrammatically shown in the right part of Fig.~\ref{f3}.  In this
study we employ the fixed center approximation\cite{Bayar:2011qj} to
calculate this multiple scattering amplitude, where the kaon
absorption width is effectively taken into account in the kaon
propagator so as to reproduce the result of the width of the $\bar{K}
N N$ bound state in Ref.\cite{Bayar:2012hn}.  In the present
formulation and parameters, the multiple kaon scattering amplitudes in
the fixed center approximation generate a peak at around $2350 \mev$
together with a resonance pole $2354 - 36 i \mev$.

\begin{figure}[b]
  \centering
  \includegraphics[width=7.5cm]{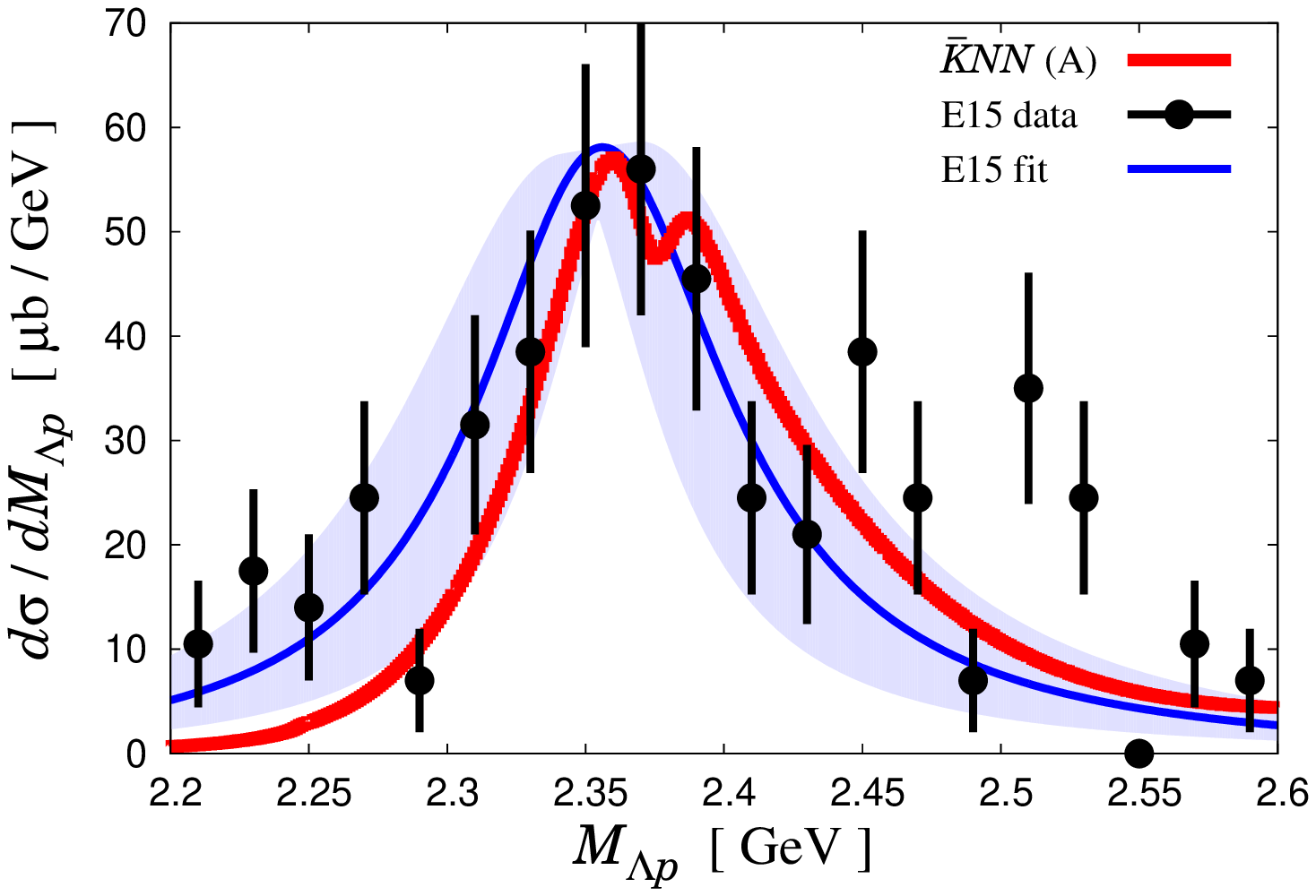} ~
  \includegraphics[width=7.5cm]{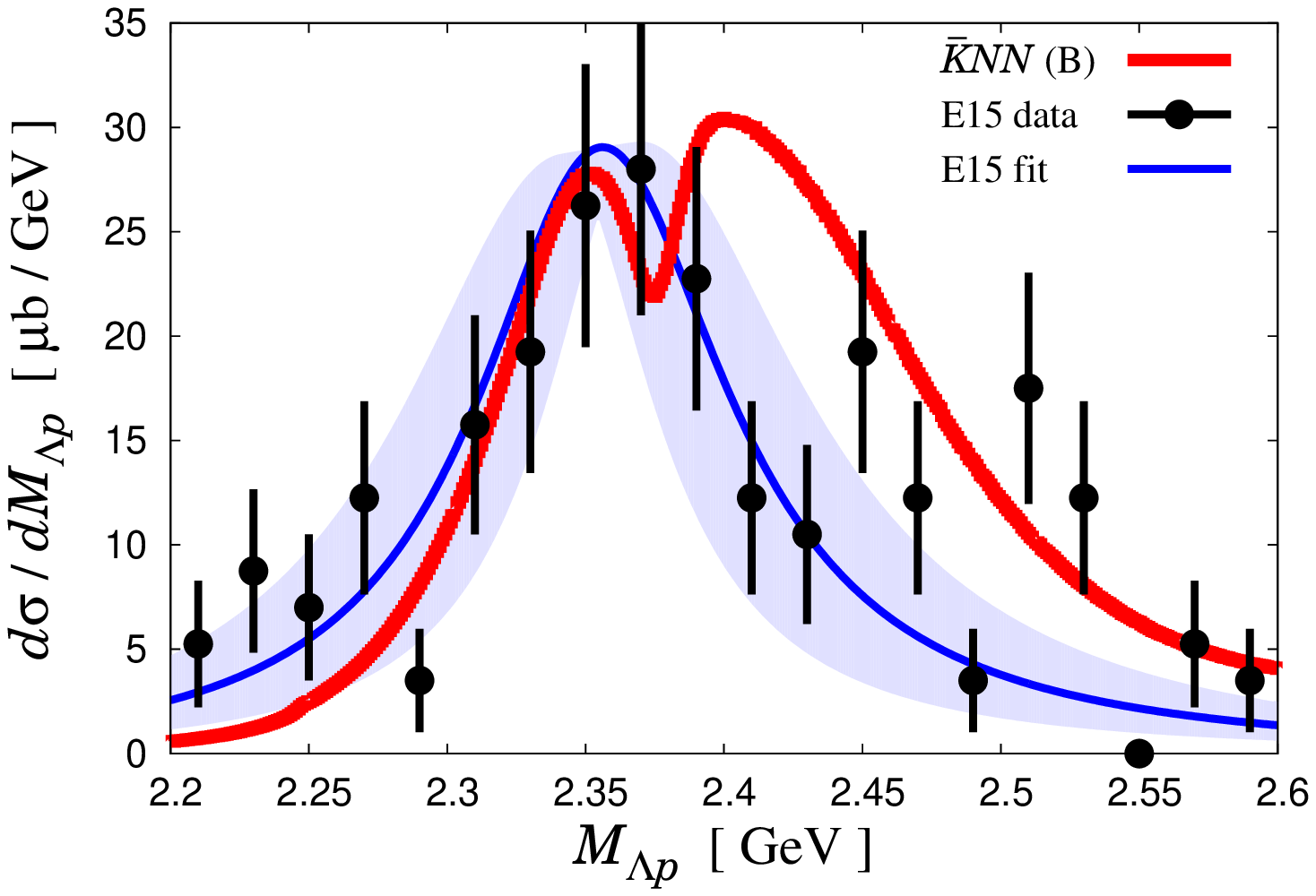}
  \caption{The same as Fig.~\ref{f2} but in the $\bar{K} N N$ bound
    scenario\cite{Sekihara:2016vyd}.  Calculations are done in option
    A (left) and option B (right).}
\label{f4}
\end{figure}

Now let us calculate the $\Lambda p$ invariant mass spectrum in the
$\bar{K} N N$ bound scenario.  The result is shown in Fig.~\ref{f4}
together with the experimental (E15) data and its
fit\cite{Sada:2016vkt} in arbitrary units.  An important finding is
that our mass spectrum is consistent with the experimental one within
the present error.  In particular, we can reproduce the tail at the
lower energy $\sim 2.3 \gev$.  This implies that our spectrum supports
the explanation that the E15 signal in the $\HeT (K^{-}, \Lambda p) n$
reaction is indeed a signal of the $\bar{K} N N$ bound state.  In
addition, we observe a two-peak structure below and above the $\bar{K}
N N$ threshold.  Actually, the lower peak is the signal of the
$\bar{K} N N$ bound state, and the higher peak comes from the
quasi-elastic kaon scattering in the first step, i.e., when the kaon
after the kaon after the collision $T_{1}$ goes almost on its mass
shell.  Here we also note that the total cross section becomes $7.6
\microB$ ($5.6 \microB$) in option A (B), which is qualitatively
consistent with the empirical value $7 \pm 1
\microB$\cite{Sada:2016vkt}.

\begin{figure}[t]
  \centering
  \includegraphics[width=7.5cm]{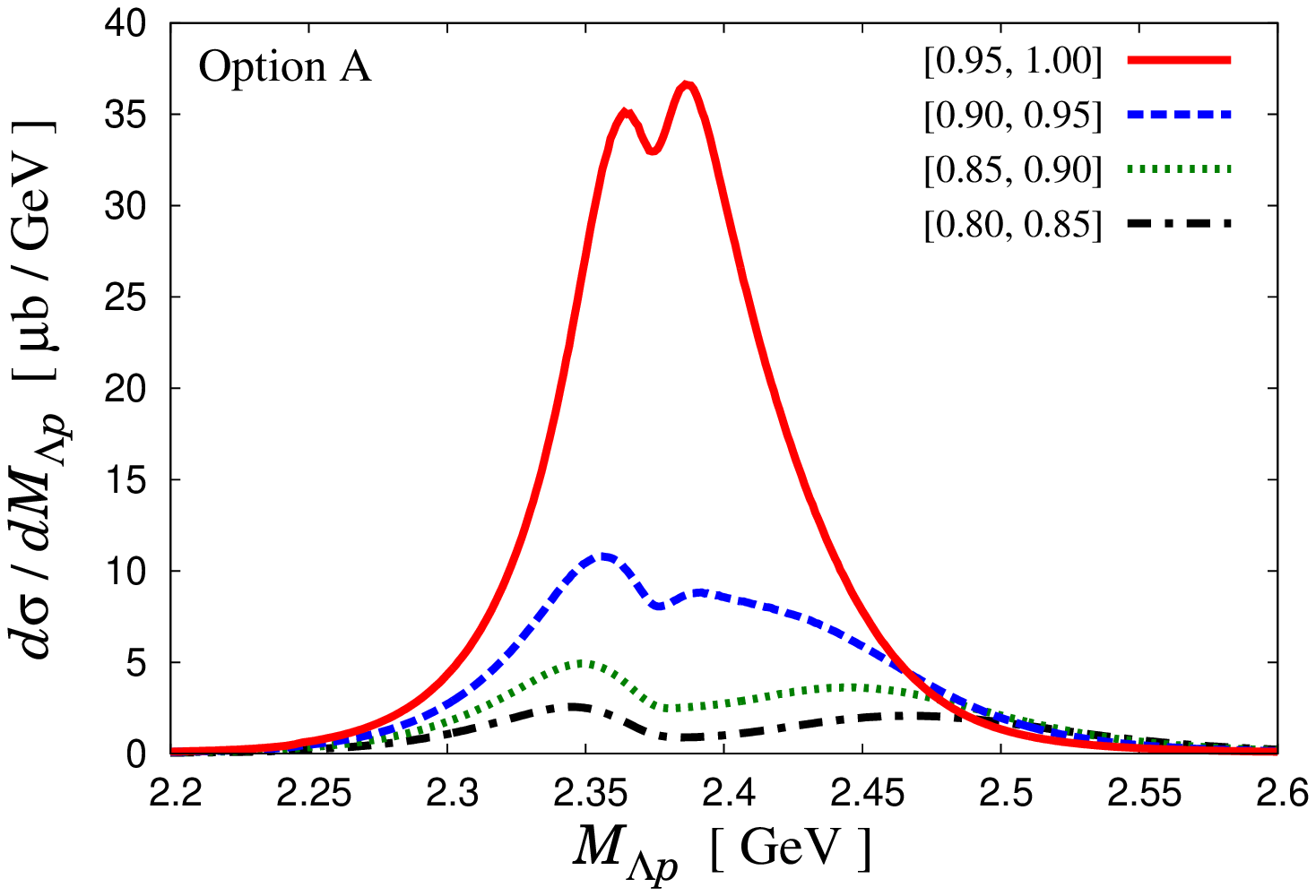} ~
  \includegraphics[width=7.5cm]{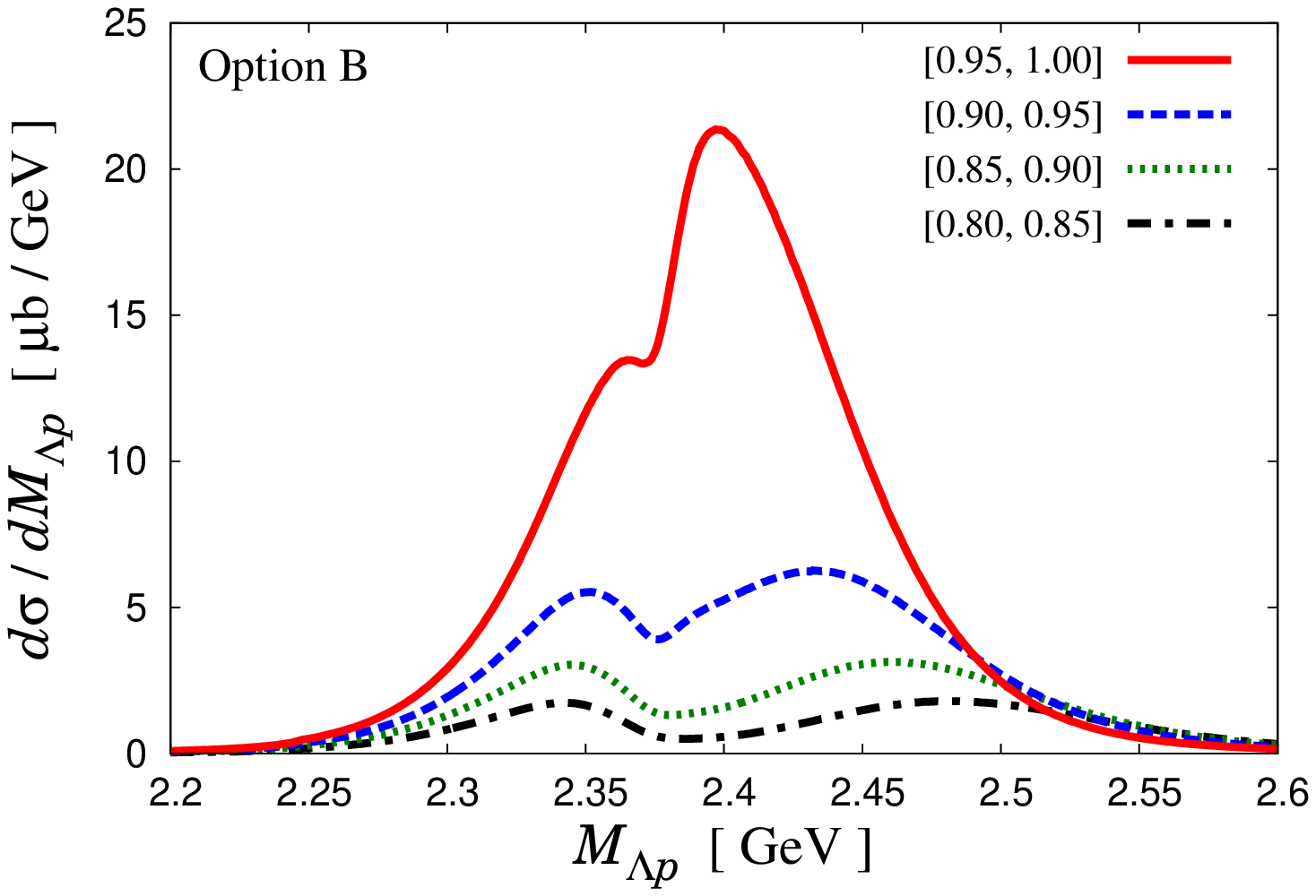}
  \caption{Mass spectrum in the $\bar{K} N N$ bound scenario with the
    angular restriction $\cos \theta _{n}^{\rm cm} \in$ [0.95, 1.00],
    [0.90, 0.95], [0.85, 0.90], and [0.80, 0.85].  Calculations are
    done in option A (left) and option B (right).}
\label{f5}
\end{figure}

Finally, in Fig.~\ref{f5} we show the mass spectrum with the
restriction of the final-state neutron scattering angle: $\cos \theta
_{n}^{\rm cm} \in$ [0.95, 1.00], [0.90, 0.95], [0.85, 0.90], and
[0.80, 0.85].  From the figure, one can see that the mass spectrum is
dominant in the forward neutron scattering, $\cos \theta _{n}^{\rm cm}
= 1$, and is suppressed as the scattering angle $\theta _{n}^{\rm cm}$
increases.  Furthermore, we can observe how the two peaks move as the
scattering angle $\theta _{n}^{\rm cm}$ increases; the signal of the
$\bar{K} N N$ bound state stays at $\sim 2.35 \gev$, but the peak of
the quasi-elastic kaon scattering shifts upward due to its kinematics.

\section{Summary and outlook}

In this study we have investigated the origin of the peak structure
near the $\bar{K} N N$ threshold in the $\HeT (K^{-} , \Lambda p) n$
reaction observed in J-PARC E15 experiment.  From the calculation of
the $\Lambda p$ invariant mass spectrum, the experimental signal is
qualitatively well reproduced in the scenario that the $\bar{K} N N$
bound state is indeed generated, and we can definitely discard the
scenario that the $\Lambda (1405)$ is generated but does not correlate
with the $p$.

The final goal of the present study is to prove that the E15 peak is
indeed the $\bar{K} N N$ signal with the help of experimental studies.
In order to achieve this, we need to check consistency between
experiments and theories for various quantities.  In this line, much
is expected from the high statistics data that are coming from the
second run of E15\cite{E15-2}.  With more data, we will be able to
compare more things such as the angular dependence of the signal and
the $\Lambda p / \Sigma ^{0} p$ branching ratio\cite{Sekihara:2009yk,
  Sekihara:2012wj}.

\end{document}